\documentstyle[prl,aps,preprint]{revtex}
\begin{document}
\title{See-Sawless Neutrino Masses} 
\author{P. Q. Hung}
\address{Dept. of Physics, University of Virginia, Charlottesville,
Virginia 22901}
\date{\today}
\maketitle
\begin{abstract}
An alternative to the famous see-saw mechanism is proposed to explain
the smallness of the neutrino masses (if present). This model involves
a fourth family which mixes {\em very little}
with the other three. It contains one heavy neutrino ($m_N > m_Z /2$)
and three very light neutrinos whose masses are radiatively induced. 
In contrast with the see-saw mechanism, all neutrino masses are {\em Dirac 
masses}. In one particular scenario, the three light neutrinos are almost
degenerate in mass and are found to be consistent with fits to the Solar and 
Atmospheric neutrino deficits. They might even account for the Hot Dark
Matter.
\end{abstract}


The possible presence of small neutrino masses and of neutrino
oscillations is believed to be a plausible explanation for
a set of experimental ``discrepancies'' and ``evidences'':
the solar neutrino problem, the atmospheric
neutrino problem, and the Liquid Scintillation Neutrino Detector (LSND)
data\cite{mohapatra}. These experimental results, naturally, will have to be 
confirmed in the future.

In this paper, we propose an alternative way of looking at the 
neutrino mass problem without 
resorting to the famous see-saw mechanism\cite{seesaw}. The most
important difference with the see-saw mechanism is the
fact that, in our model, the neutrinos have {\em only}
Dirac masses. 

We now list our three main assumptions.

I) There is a {\em non-sequential} fourth family. By
non-sequential, we mean that this fourth family is isolated from
the first three families by tiny mixing angles.  One way to realize this picture is to
assume an almost unbroken 3 + 1 structure under a ``light'' horizontal
family symmetry to isolate the fourth family. By ``light'' horizontal family symmetry,
we mean a symmetry {\em among the first three families}.

Recently, it was found\cite{hung} that such a fourth generation with a
quark mass $\sim$ 150 GeV helps bring about a unification of the
SM gauge couplings at a scale $\sim 3.5 \times 10^{15}$ GeV, corresponding
to a partial proton lifetime $\sim 3.3 \times 10^{34 \pm 2}$ years, in
a non-supersymmetric $SU(5)$ model. In Ref. \cite{frampton}, a search was proposed
for long-lived quarks which can arise in such a model.

II) There is {\em one and only one} right-handed neutrino, $N_R$, which is
a singlet under that horizontal symmetry. (This is in contrast with
the usual picture where there is one right-handed neutrino for
each species.)

III) There is an exact global L (lepton number) symmetry and that
$N_R$ carries the same lepton number as all the other leptons. (In the context
of simple grand unification,
it would be the B-L global symmetry of SU(5)
with minimal Higgs.) This symmetry
would forbid a Majorana mass term of the form $N_R N_R$ for $N_R$. If
one is not concerned about any grand unification, then global lepton
number conservation alone is sufficient to forbid such a term.


The Yukawa couplings 
which respect the ``light'' horizontal symmetry
(The SM Higgs field $\phi$ is assumed to be a singlet under that
symmetry), are of the form:
1) $G_{l} (\bar{l}_{L}^{i} \phi e_{R,i} + h. c.)$;
2) $G_{E} (\bar{L}_{L} \phi E_{R} +h. c.)$;
3) $G_{N} (\bar{L}_{L} \tilde{\phi} N_{R} +h. c.)$,
where $\tilde{\phi} = i \sigma_2 \phi^{\ast}$, $i=1,2,3$ (
the ``light'' indices), and $L_L = (N_L, E_L)$.
When $<\phi> = (0, v/\sqrt{2})$, the $4 \times
4$ charged lepton mass matrix is {\em diagonal} and the only neutral lepton
that gets a (Dirac) mass is the fourth family $N$. At this level, 
there is {\em no} mixing between $E$ and the light leptons. 
It is thus natural, at tree level,
to have a massive Dirac fourth neutrino and three massless neutrinos. 
Assumption (III) (L or B-L symmetry) forbids a Majorana mass term $N_R N_R$.

To proceed further, one needs to embed the (light) horizontal
symmetry into a larger one. For this purpose, 
let us assume the ``light'' family symmetry to be described by 
the group $SO(3)$ with the first three families transforming
as a 3-dimensional vector representation and with the fourth family being
a singlet. Let us now assume that there is a Grand Family gauge symmetry group
and it is $SO(4)$. We shall choose the following
basis for the $SO(4)$ generators: $M_{ij}$ and $M_{4i}$, with $i,j = 1,2,3$, which
are the generators of the $SO(3)$ subgroup and $SO(4)/SO(3)$ factor group
respectively. The gauge bosons of $SO(3)$ couple to $M_{ij}$ while
those of $SO(4)/SO(3)$ couple to $M_{4i}$.
When $SO(4)$ breaks down to $SO(3)$, it is this
$SO(3)$ that we identify with the ``light'' family symmetry.
Under $SO(4)$, the left and right-handed leptons would transform
as $\psi^{\alpha}_L = (l_{L}^i , L_{L})$ and $e^{\alpha}_R =
( e_{R}^i , E_R)$, where the superscripts $\alpha= 1,..,4$ and
$i = 1, 2, 3$ denote the $SO(4)$ and   ``light''
lepton family indices respectively. Since, in our model, there is only one right-
handed neutral lepton, $N_R$, it would automatically be a {\em singlet}
under the Grand Family Symmetry $SO(4)$.

Under $SO(4)$, the only invariant Yukawa coupling  that can be 
written is $G_{L} \bar{\psi}^{\alpha}_{L} \phi
e_{R,\alpha}$. This alone would be unsatisfactory from a phenomenological
viewpoint since it would give equal masses, $m^{0}_E$,
to all four charged leptons. Furthermore, from LEP2, one has $m_E > m_W$ and
since the fourth family is assumed to mix very little with the other
three , one cannot use some kind
of democratic mass matrix (with unity everywhere) to make $E$ much
heavier than the lighter three. An extra term is needed to give an
additional mass to the fourth charged lepton. 

Another important question concerns the nature of the Yukawa term  $G_{N} 
(\bar{L}_{L} \tilde{\phi} N_{R} +h. c.)$ which
gives a Dirac mass to the ``heavy'' 4th
neutrino $N$. It is, however, not $SO(4)$ invariant (although
it is invariant under the ``light'' family symmetry $SO(3)$). To be consistent,
this Yukawa coupling should be derived from an $SO(4)$-invariant term . 

To address the above issues, let us introduce the Higgs fields
needed to break $SO(4)$. For instance, to break $SO(4)$ completely, one might
use four Higgs fields belonging each to a vector representation. It 
is beyond the scope of this paper to discuss the details of such a breaking and we
shall assume that it can be done. Let us call one of such 4-dimensional
Higgs fields $\Omega$ where
$\Omega = (\Sigma^{i}, \Theta)$, with $i=1,2,3$.
Let us assume that $\Omega$ develops the following
vacuum expectation value (VEV): $<\Omega> = (0, <\Theta>)$, with $<\Theta> =
{\cal M}$ being typically the scale of $SO(4)$ breaking. In fact,
if $\Omega$ were the only Higgs field present for $SO(4)$, its VEV
would spontaneously break $SO(4)$ down to the ``light'' family symmetry
$SO(3)$.

We now propose the following minimal
set of extra ``superheavy'' fermions- singlets under $SO(4)$-
whose attractive feature is
to generate tree-level masses for $N$ and $E$. These are the
fermions which can couple to $\Omega$. They are (under $SU(2)_L 
\otimes U(1)_Y$):
1) $F_{L,R} = (2, -1/2)$;
2) $M_{L,R} = (1, -1)$.
These extra fermions are vector-like under the SM. As a result, they can
have the following gauge-invariant mass terms: ${\cal M}_{F}\bar{F}_{L} 
F_{R} + h.c.$ and ${\cal M}_{M} \bar{M}_{L} M_{R} + h.c.$, where
${\cal M}_{F}$ and ${\cal M}_{M}$ are {\em assumed} to be of the order of
the $SO(4)$ breaking scale. We propose the following Yukawa interactions
which respect $SO(4) \otimes SU(2)_{L} \otimes U(1)_Y$:
\begin{eqnarray}
{\cal L}_Y& =& G_1 \bar{\psi}^{\alpha}_{L} \Omega_{\alpha} F_{R} +
G_2 \bar{F}_{L} \tilde{\phi} N_{R} +
G_3 \bar{F}_{L} \phi M_{R} + \nonumber \\
          &  &G_4 \bar{M}_{L} \Omega_{\alpha} e^{\alpha}_{R},
\end{eqnarray}
where $\alpha= 1,..,4$ is the $SO(4)$ index. We shall endow $\psi^{\alpha}_L$,
$e^{\alpha}_R$ and $\Omega$ with a discrete symmetry so that $\psi^{\alpha}_L$
and $e^{\alpha}_R$ couple {\em only} to $\Omega$
and not to other 4-dimensional $SO(4)$ Higgs fields.
(This discrete symmetry could be, for instance, the simultaneous change of
sign of these three fields.)

Integrating out the heavy fields
$F$ and $M$ below the $SO(4)$ breaking scale and
with $<\Omega> = (0, <\Theta>)$, 
it is straigthforward to derive the following effective
Yukawa terms: $G_N \bar{L}_L \tilde{\phi} N_R$ with
$G_N = G_1 G_2 <\Theta> / {\cal M}_F$, and
$G_E \bar{L}_L \phi E_R$ with $G_E = G_1 G_3 G_4 <\Theta>^{2}/ {\cal
M}_F {\cal M}_M$. From these terms we obtain the following
masses: $m_{0N} = G_1 G_2 \frac{<\Theta>}{{\cal M}_F} \frac{v}{\sqrt{2}}$, and
$\tilde{m}_E = G_1 G_3 G_4 \frac{<\Theta>^{2}}{{\cal M}_F {\cal M}_M} \frac{v}{\sqrt{2}}$,
where $<\Theta> \sim {\cal M}_F$.
The total mass of the fourth charged lepton, $m_E$, would be the sum of $\tilde{m}_E$
and $m^{0}_E$ (the mass which is common to all four charged leptons). 
Phenomenologically, one could have $m^{0}_E \ll \tilde{m}_E$ which would
provide the desired hierarchy.

There are two steps that one could do to compute the ``light''
neutrino masses. These steps are depicted in Figs.1 and 2 which
show the $E-e^{i}$ mixing and the
effective Yukawa term $ \bar{\nu}^{i}_L \phi^{0} N_R$ respectively.
We shall assume that $SO(3)$ breaking will
give rise to a non-diagonal mass matrix for the ``light'' charged lepton sector.
(Its detailed form and mechanism is not essential to the arguments
presented below.) 

In Fig.1, we have introduced another quartet of Higgs field
which we denote by $\tilde{\Omega} = (\tilde{\Sigma}^{i}, \tilde{\Theta})$.
(This is one of several Higgs field needed to spontaneously
break $SO(4)$.) First, this particular quartet is prevented
by the previous discrete symmetry from coupling to the fermions.
Secondly, in contrast with $\Omega$, we require that 
$<\tilde{\Omega}> = (<\tilde{\Sigma}^{i}>, <\tilde{\Theta}>)$
so that one obtains an effective Yukawa coupling of the form:
$C_{i} \bar{e}^{i}_L \phi^{0} E_R$, where $C_i$ is a constant containing
the loop inegration and various couplings. (The contribution of
$\Omega$ to $C_i$ is zero because $<\Omega> = (0, <\Theta>)$.)
This can be easily seen because such a Yukawa mixing between
$e^{i}_L$ and $E_R$ can only arise when $SO(3)$ is itself broken.
This kind of vacuum expectation value can be arranged in a general
potential. It is beyond the scope of the paper to present it here.
We shall assume that it can be done.

The gauge bosons in Fig.1 are the massive $SO(4)/SO(3)$ gauge
bosons. Without loss of generality, we shall assume that their
masses, $M_4$, are of the order of $<\tilde{\Theta}>$. With this
in mind, the coefficient $C_i$ can be computed to be:
$C_i = \tilde{\lambda} (\frac{g_F^2}{16 \pi^2})(\frac{\sqrt{2}
m_{e_i}}{v}) \frac{<\tilde{\Sigma}^i>}{M_4} \ln (\frac{M_4^2 + m_{\tilde{\Theta}}^2}
{m_{\tilde{\Theta}}^2})$,
where $g_F$, $m_{\tilde{\Theta}}$, and $M_4$ are the $SO(4)$ gauge coupling, 
the mass of $\tilde{\Theta}$, and the mass of the $SO(4)/SO(3)$ gauge
bosons respectively. The factor $\tilde{\lambda}$ comes from the cross
coupling $\tilde{\lambda} \phi^{\dag} \phi \tilde{\Omega}^{\dag}
\tilde{\Omega}$. Also, in $C_i$, $m_{e_i}$ is the mass eigenvalue
of the charged lepton $e_i$.  We shall comment below on the
possible ranges for the various parameters in $C_i$.

In the following discussion, for simplicity, we shall assume
that $<\tilde{\Sigma}^{i}> = M_3$, independent of $i$.
$M_3$ will be related to the scale
of $SO(3)$ breaking. 

Using $C_i \bar{e}^{i}_L \phi^{0} E_R$, one can now calculate 
(in the Feynman-'t Hooft gauge) the
diagrams shown in Fig.2a,b. The various factors that enter the vertices
of the diagrams need some explanations. For $\bar{E}_L \phi^{-} N_R$, one has
$\sqrt{2} m_{0N}/ v$. $C_{i} \bar{e}^{i}_L \phi^{0} E_R$ gives the 
$\bar{e}_{iL} E_R$ mixing.
The $\bar{\nu}_{iL} \phi^{+} e_{jR}$ vertex contains a term
$V_{ij} \sqrt{2} m_{e_j} / v$, where $V_{ij}$'s are actually 
elements of the matrix that diagonalizes the {\em charged} lepton
sector. The reason is that there is only {\em one} right-handed
neutral lepton, $N_R$, and the neutral lepton mass matrix is
necessarily diagonal. We show below what the effective
right-handed component for {\em each} neutrino will be.

Diagram (2a) is proportional to $m_{e_i}^2 m_E^2$ while Diagram (2b)
is proportional to $m_{e_i}^3 m_E$ with similar coefficients in
front. Therefore Diagram (2a) is larger than (2b) by a factor
$m_E / m_{e_i}$, which is much larger than unity even for
$m_{e_i} = m_{\tau}$. Because of this we list below the
dominant contribution to Fig.2, namely Fig.2a. 
We obtain:
\begin{equation}
\frac{m_{0\nu_i}}{m_{0N}} = K \sum_{j=1}^{3} V_{ij} \frac{m_{e_j}^{2}}{v^2},
\end{equation}
where
\begin{eqnarray}
K& =& \tilde{\lambda}\frac{g_{F}^2}{16 \pi^2} \frac{M_3}{M_4}
\frac{1}{4 \pi^2} \frac{m_{E}^2}{m_{W}^2} \ln (\frac{M_4^2 + m_{\tilde{\Theta}}^2}
{m_{\tilde{\Theta}}^2})\{\ln (\frac{M_{W}^2 + m_{E}^2}
{m_{E}^2}) \nonumber \\ 
 &   & + \frac{m_{E}^2}{M_{W}^2 + m_{E}^2}\}.
\end{eqnarray}

We first show how $m_{0\nu_i}$ and $m_{0N}$ are related to the Dirac masses
of the neutrinos. We then discuss their relative magnitudes.

Let us write the part of the Lagrangian containing the kinetic and
mass terms for the four neutrinos. For simplicity, we shall omit
the gauge part in the kinetic term. 
One has:  ${\cal L}^{(\nu)} = \sum_{j=1}^{3} i\bar{\nu}_{jL}\not\!\partial \nu_{jL}
+ i\bar{N}_L\not\!\partial N_L + i\bar{N}_R \\
\not\!\partial N_R
- m_{0N} (\bar{N}_L N_R + h.c.) - \sum_{j=1}^{3} m_{0i} (\bar{\nu}_{jL}
N_R + h.c.)$ .
Let us introduce the following 4-component Dirac spinors:
$\tilde{\nu}_i = (\nu_{iL} , \alpha N_R)$ ;
$\tilde{N} = (N_{L}, \alpha N_R)$.
By writing ${\cal L}^{(\nu)}$ in terms of the above Dirac spinors and
comparing it with the previous expressions, it is easy to see that 
one has to have
$\alpha = 1/2$. It means that the Dirac masses, $m_{\nu_i}$ and
$m_N$, coming from terms like $m_{\nu_i}\bar{\tilde{\nu}}_i \tilde{\nu}_i$ and
$m_N \bar{\tilde{N}} \tilde{N}$, are now given by:
$m_{\nu_i} = 2\,m_{0\nu_i}; m_N = 2\,m_{0N}$. The ratio, however,
remains the same, i.e.
$m_{\nu_i}/m_N = m_{0\nu_i}/m_{0N}$. 


To know the relative magnitude of $m_{\nu_i}$ compared with $m_{N}$,
one needs to have an estimate for various parameters which appear in $K$.
To be more specific, let us take a definite example. For instance,
let us assume that
$m_E \sim 2 m_W$, $M_3 \sim M_4$, $g_F \sim g_{weak}$.
A crude estimate then gives $K \sim 1.5 \times 10^{-3} \tilde{\lambda}$. Now
let us recall that $\tilde{\lambda}$ comes from $\tilde{\lambda} \phi^{\dag} \phi 
\tilde{\Omega}^{\dag} \tilde{\Omega}$. In order to prevent the SM Higgs
field from acquiring a large mass, it is easy to see that the
constraint on $\tilde{\lambda}$ is approximately $\tilde{\lambda} < v^2 / M_{4}^2$
where we have set the SM quartic coupling $\lambda \sim O(1)$.
In consequence, if $v \ll M_4$, $\tilde{\lambda}$ can be {\em very} small. (This
is the familiar statement of gauge hierarchy encountered in the construction
of Grand Unified theories.) For example, if the Grand Family breaking is in
the TeV region so that, say, $v^2 / M_{4}^2 \sim 10^{-4}$, one then has
$\tilde{\lambda} < 10^{-4}$ implying $K < 1.5 \times 10^{-7}$. Since 
$|V_{ij}| \lesssim 1$
and the heaviest charged lepton mass is $m_{\tau} = 1.784$ GeV, the
heaviest ``light'' neutrino can have a mass of roughly 1.3 eV for $m_N \sim
160$ GeV. It could, of course, be heavier or lighter depending on, e.g.
the ratio $v^2 / M_{4}^2$, among other things.

Let us now turn to a more complete discussion of all neutrino masses. From $C_i$,
one can see that, for the range of parameters mentioned above, the mixing
between $E$ and the ``light'' charged leptons is very small, i.e. $\sim
10^{-5}$. In consequence, the $3\times 3$ mixing submatrix, among the ``light''
charged leptons, whose elements are $V_{ij}$ will be approximately unitary.
Without loss of generality, we shall use the standard CKM parametrization
for such a matrix, neglecting any possible CP violation effect. The
weak charged current can be written, in terms of mass eigenstates, as: 
$J_{\mu} = 2 \sum_{i,j=1}^{3} \bar{l}_{Li} V_{ij} \nu_{Lj}$. Here
$l_{Li} = (e, \mu, \tau)$ and $\nu_{Lj}$ are the mass eigenstates. Also
$V_{ij} = U_{l}^{\dagger} U_{\nu}$, where $U_{l}$ and $U_{\nu}$
are the matrices which diagonalize the charged and neutral lepton
sectors respectively. As we have stated earlier, the neutral lepton
mass matrix is diagonal because there is only one right-handed
neitrino. As a result, $U_{\nu} = 1$ and $V_{ij} = U_{l}^{\dagger}$.
The neutrino masses computed below are {\em directly} related to
the matrix that diagonalizes the charged lepton sector. The ``light''
neutrino mass eigenvalues can now be written in terms of the ``light'
charged lepton mass eigenvalues as:
\begin{eqnarray}
m_{\nu_1}& =& \frac{K m_N}{v^2} \{ (c_{\omega} c_{\phi})m_{e}^2 + (-s_{\omega}
c_{\psi}-c_{\omega}s_{\psi}s_{\phi})m_{\mu}^2 + \nonumber \\ 
         &  &(s_{\omega}s_{\psi}-c_{\omega}c_{\psi}s_{\phi})m_{\tau}^2 \} \\
m_{\nu_2}& =& \frac{K m_N}{v^2} \{ (s_{\omega} c_{\phi})m_{e}^2 + (c_{\omega}
c_{\psi}-s_{\omega}s_{\psi}s_{\phi})m_{\mu}^2 + \nonumber \\ 
         &  &(-c_{\omega}s_{\psi}-s_{\omega}c_{\psi}s_{\phi})m_{\tau}^2 \} \\
m_{\nu_3}& =& \frac{K m_N}{v^2} \{ s_{\phi}m_{e}^2 + s_{\psi}c_{\phi}m_{\mu}^2 + 
c_{\psi}c_{\phi}m_{\tau}^2 \} ,
\end{eqnarray}
where $c\equiv cos$ and $s \equiv sin$. The angles $\omega, \phi, \psi$ are used in Ref.
\cite{kuo}. The hierarchy, if any, among the $m_{\nu_i}$'s depends on several factors:
the relative magnitudes of the charged lepton masses and the magnitude of the
coefficients appearing in front of these masses.
In the discussion of neutrino oscillation, the angles that appear there come
from the equation that expresses the flavour eigenstates in terms of the
mass eigenstates, namely $\nu^{0}_{Li} = \sum_{j=1}^{3} V_{i j} \nu_{Lj}$,
where $\nu_{Li}^{0} = (\nu_{e}, \nu_{\mu}, \nu_{\tau})$.
Therefore, the neutrino masses are intrinsically linked
to the oscillation angles. Next,
one observes that: $m_{\mu}^2 / m_{\tau}^2 \sim 3.5072 \times 10^{-3}$ and
$m_{e}^2 / m_{\tau}^2 \sim 10^{-7}$, $m_{e}^2 / m_{\mu}^2 \sim 2.3 \times 10^{-5}$.
Unless the coefficients appearing in front of $m_{\mu}^2$ and $m_{\tau}^2$ are
extremely small- a rather unusual scenario- it is safe to say that the
main contributions to the masses come from $m_{\mu}^2$ and $m_{\tau}^2$ in
Eqs. (4,5,6). We thus neglect the $m_{e}^2$ terms from hereon.

The mass differences
relevant to neutrino oscillation are customarily written as:
$\Delta m_{ij}^2 = |m_{\nu_i}^2 - m_{\nu_j}^2|$. From Eqs. (4,5,6), one
can compute the following relevant ratios:
$\frac{\Delta m_{12}^2}{m_{\nu_3}^2}$ and
$\frac{\Delta m_{23}^2}{m_{\nu_3}^2}$. These ratios depend only on the three
angles and on the known masses, $m_{\mu}$ and $m_{\tau}$. The actual
values of $\Delta m_{ij}^2$ will also depend on $m_{\nu_3}$ which is
a free parameter since it depends on various factors such as,
for example, the breaking scale of the Grand Family symmetry, among others.
Its magnitude could have interesting
implications concerning the physics of the family symmetry. 

It is beyond the scope of this paper to do a detailed analysis of various
possible scenarios using Eqs. (4,5,6). We choose to illustrate the predictive
power of these equations by taking the following example. We ask the
question: Given $m_{\nu_3}$ (input) and the three angles $\phi$, $\omega$ and
$\psi$, would one obtain consistent values for $\Delta m_{ij}^2$ and the oscillation
angles? What would that imply as far as the three neutrino masses are
concerned? After a quick scan of various angles, we choose:
$\phi = 2.9^{0}$, $\omega = 47.04325^{0}$, $\psi = 55.01^{0}$, and
$m_{\nu_3}^2 = 2.5\, eV^2$. Plugging these values into the ratios
$\frac{\Delta m_{12}^2}{m_{\nu_3}^2}$ and
$\frac{\Delta m_{23}^2}{m_{\nu_3}^2}$, we obtain:
$\Delta m_{12}^2 = 8.65 \times 10^{-6} eV^2$; $\Delta m_{23}^2 = 0.0233 eV^2$;
$\sin^2 2\theta_{e\mu} = 4 s_{\phi}^2 c_{\phi}^2 s_{\psi}^2 =\sin^2 2\phi s_{\psi}^2
=6.85 \times 10^{-3}$; and $\sin^2 2\theta_{\mu\tau} = 4 s_{\psi}^2 c_{\psi}^2 
c_{\phi}^4 =\sin^2 2\psi c_{\phi}^4 = 0.878$. These values are consistent
with the small-angle MSW solution to the Solar neutrino deficit, and
with the sub-GeV and multi-GeV data for the solution to the Atmospheric
neutrino deficit\cite{mohapatra}. Furthermore, from the 
relationship between the neutrino
masses, we deduce the values for $m_{\nu_1}$ and $m_{\nu_2}$, given the
above values of the angles and of $m_{\nu_3} = 1.5811388\, eV$. We get:
$m_{\nu_1} = 1.5884876\, eV$ and $m_{\nu_2} = 1.5884904\, eV$. Interestingly,
one has here a case of almost degenerate neutrinos: $m_{\nu_1} \approx
m_{\nu_2} \approx m_{\nu_3}$. Notice that $m_{\nu_1} + m_{\nu_2} +
m_{\nu_3} \approx 4.78\, eV$, which seems to be a preferred value
for the hot dark matter in a mixed hot-cold scenario\cite{mohapatra,babu}. 
The neutrino masses
in our scenario are of a Dirac nature and, as a consequence, they
should {\em not} give rise to neutrinoless double beta decay. Their masses
can be $\sim$ 1.6 eV and are not subject to the upper limit (for
a Majorana mass) of $\sim$
0.56 eV from the search for neutrinoless double beta decay by the Heidelberg-
Moscow $^{76}Ge$ experiment\cite{klapdor}. Note, in passing, that, for the values of 
$\Delta m_{ij}^2$ presented above, even if $m_{\nu_i} \sim 1.6\, eV$, there
appears to be no problem with the so-called $r$-process in supernova
nucleosynthesis of heavy elements. One last (but not least) remark:
with $m_{\nu_3} \sim 1.6$ eV and assuming, e.g. $m_N \sim 160$ GeV, we
obtain $M_4 \sim M_3 \sim 18$ TeV, where the expression for $K$ has been
used. The Family scales $M_{3,4}$ can vary, depending on a number of
factors contained in $K$. If $m_{\nu_3}$ were to be much smaller,these
scales will be correspondingly much larger than the previous rough
estimate. Nevertheless it is interesting to see, in this scenario, the
deep connection between neutrino masses and Family scales.

I would like to thank Qaisar Shafi for an insightful comment.
This work is supported in parts by the US Department
of Energy under grant No. DE-A505-89ER40518.

\begin{figure}
\caption{Diagram showing the computation of the effective Yukawa term:
$C_{i} \bar{e}^{i}_L \phi^{0} E_R$.}
\end{figure}
\begin{figure}
\caption{Diagrams showing the computation of the effective Yukawa term:
$ \bar{\nu}^{i}_L \phi^{0} N_R$, in the Feynman-'t Hooft gauge.
The large dot between $e^{i}_L$ and $E_R$ represents
the mixing computed from the diagram shown in Fig. 1}
\end{figure}
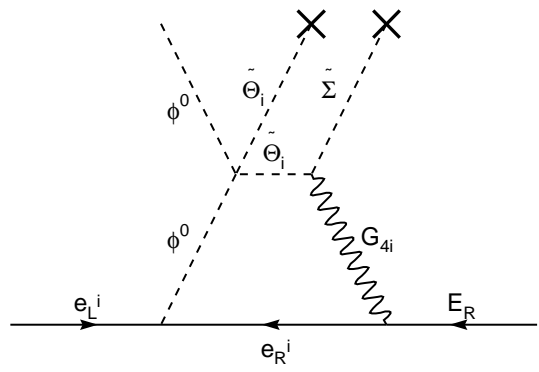
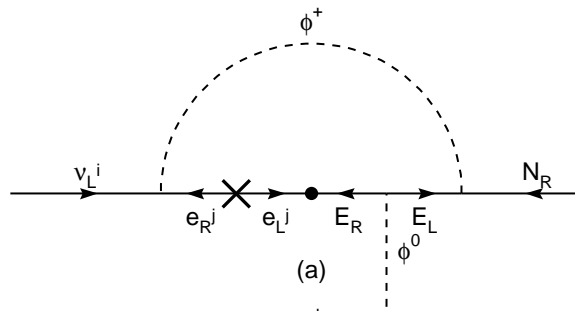
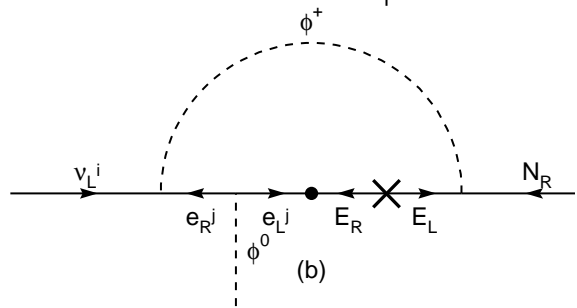
\end{document}